# Coriolis force-based instability of a shear-thinning microchannel flow

Saunak Sengupta[1], Sukhendu Ghosh[2], and Suman Chakraborty[1]

[1]Department of Mechanical Engineering, Indian Institute of Technology Kharagpur, Kharagpur-721302, India

[2]Department of Mathematics, Indian Institute of Technology, Jodhpur-342037, India

**Abstract**

Instability mechanism based on Coriolis force, on a rapidly rotating portable device handling shear-thinning fluids such as blood, is of utmost importance for eventual detection of diseases by mixing with the suitable reagents. Motivated by this proposition, the present study renders a modal stability analysis of shear-thinning fluids in a rotating microchannel modelled by the Carreau rheological law. When a microchannel is engraved a rotating compact disc (CD)-based device, the centrifugal force acts as the driving force that actuates the flow and the Coriolis force enhances the mixing process in significantly short span by destabilizing the flow. An Orr-Sommerfeld-Squire analysis is performed to explore the role of these forces on the linear stability of rotating shear-thinning flow. Reported results on shear-thinning flow with streamwise disturbances indicate that the critical Reynolds number for the flow transition with viscosity perturbation is nearly half of that of the critical value for the same without viscosity perturbation. In sharp contrast, the present analysis considering spanwise disturbances reveals that the critical Reynolds numbers with and without viscosity perturbation remain virtually unaltered under rotational effects. However, the viscosity variation has no significant influence on the Coriolis force-based instability. Numerical results confirm that a momentous destabilization is possible by aid of the Coriolis force via generating secondary flow inside the channel. Interestingly, the roll cells corresponding to the instabilities at lower time constants exhibit the existence of two distinct vortices, and the centre of the stronger one is essentially settled towards the unstable "stratified" region. Moreover, for a higher value of the time constant, only one vortex occupies the entire channel. This in turn, may turn out to be of fundamental importance in realizing new instability regimes facilitating efficient mixing in rotationally actuated fluidic devices deployed for biochemical analysis and medical diagnostics.

## 1. Introduction

Medical diagnosis is a procedure to determine a person's health condition based on various tests. Presently, numerous tests and diagnosis procedures are performed to assess the cause of deteriorating health conditions of a patient. These procedures can be based on invasive or non-invasive investigation including blood tests. Some applications and tests necessitate proper mixing of reagents with the blood samples, which is a challenging task and usually achieved by centrifugation. Centrifuge is an elaborate rotational device that exploits forces inherent to a rotational platform.[5] There has been a growing trend of making this procedure miniaturized, cost effective and portable. This may be achieved on a revolving compact disc (CD) based device akin to the ones used in hard drives of computers.[4] This, coupled with the facilitation of mixing of reagents mediated by rotationally induced instabilities, has recently opened up new vistas of medical diagnostics through the generation of colorimetric signals. Yet, a major challenge that remains to be addressed is to arrive at fundamental scientific principles elucidating the inception of various instability regimes triggered by microfluidic transport of complex body fluids on the spinning device, facilitating rapid diagnostics of diseases.

Microfluidics of rotational devices[1,6–9] is an emerging technology in which flow actuates due to centrifugal force as the device rotates. Such a device does not require an external pump to trigger the flow and does not necessitate elaborate fabrication procedures for manufacturing. In traditional stationary microfluidic devices, mixing occurs predominantly due to diffusion, posing significant challenges towards achieving a homogeneous solution.[10] However, in a rotating platform, enhanced mixing becomes possible by means of Coriolis force,[4,11] which generates secondary flow into the rotating system (see figure 1). On introducing secondary flow caused by Coriolis forces, an enhanced level of mixing at such a small scale, thus, becomes achievable.[12–16] This results in various advantages: (a) different types of fluid specimen can be used with different fluid properties such as, viscosity, conductivity etc., [17–22] (b) easier micro-fabrication process,[23,24] and (c) simple flow actuation mechanism ideal for implementation in resource-constrained settings.

In order to facilitate mixing processes by triggering rotational instabilities, various simple configurations have been devised, where the Coriolis force overwhelms the centrifugal force.[1,11,25–27] This has led to the realization of important biological operations, for example, separation of blood constituents from whole blood.[3,28–31] However, the underlying procedure can be made more rapid and efficient only by developing a more comprehensive understanding of the mechanism of instability under various rotational forces on a revolving CD based device, amidst rheologically complex constitution of the fluid medium.

In the recent past, increasing attention has been devoted on fluids that do not conform to the Newtonian postulate of linear relationship between stress and strain rate in simple shear, and are of significant importance in the industrial, medical, chemical disciplines. Those fluids are accordingly categorized as non-Newtonian, non-linear, complex or rheologically complex fluids. The apparent viscosity, defined as the ratio of shear stress acting on fluid and rate of shear strain, is not constant for the above category of fluids and depends either on shear stress or strain rate.[32,33] It is convenient to classify such variety of fluids (where non-Newtonian aspect is dominant) into three broad categories: (a) purely viscous, inelastic, time-independent or generalized Newtonian fluids (GNF); (b) time-dependent fluids, and (c) visco-elastic or elastico-viscous fluids.[34] In the present study, we shall only concentrate on GNF. Depending on steady shear behavior, such fluids can be further divided into three sub-categories: (i) shear-thinning or pseudoplastic behavior, (ii) visco-plastic behavior with or without shear-thinning behavior and (iii) shear-thickening or dilatant behavior. Shear-thinning fluids are perhaps the most widely encountered type of non-Newtonian fluid in engineering (like polymer melts and blends, foams or emulsions or suspensions) or medical (such as blood) practices. The apparent viscosity gradually decreases with increasing shear stress for a shear-thinning fluid.

Blood rheology is very complex in nature. Currently, no universal mathematical model is available that may represent the viscous property of blood. However, general classifications of non-Newtonian models are available in the literature, which are useful to simulate the blood flow with different degrees of accuracy. Different non-Newtonian blood models have been used previously, particularly to simulate blood flow in human subjects.[35–39]

The principal cause of non-Newtonian behavior of blood mainly results from shear-thinning caused by rouleaux clustering.[16,40,41] For an exhaustive description of the hemodynamic phenomenon, a viscous model, appropriately capturing the low and high shear rate behavior of blood, is required. The popularly used models, viz. the power law, Casson and Carreau model, have been successfully used to explain the rheological behavior of blood in many ways.[41–45] Under certain scenarios, however, due to the disaggregation and deformation of red blood cells (RBCs), the rheology of blood demonstrates viscoelastic behavior.[46–50] Nevertheless, in the present study, we have only considered the viscous and steady-state nature of flow (the elastic nature of blood and the particle size of RBCs, white blood cells (WBCs) and platelets are neglected as a first approximation). This elementary simplification makes the present analysis computationally more tractable, while providing a meaningful and relevant description of the system dynamics. [35,51–53]

Stability analysis of a micro-confined shear-thinning fluid on a revolving CD based device can help addressing many unanswered questions in the field of medical diagnostics and can open up a prospective field for future study. For example, the effect of small span-wise disturbances in presence of Coriolis force amidst complex blood rheology remains unaddressed. The importance of base flow and velocity gradient on the instability regimes of such typical shear-thinning fluids also remains elusive. In tandem, the pertinent critical parameters like shear-thinning index ($n$) and time constant ($\lambda$) for shear-thinning flow and their dependence on blood rheology remain unanswered.

Here, we bring out unique aspects of hydrodynamic instability triggered by the confluence of rotational forces and shear-thinning rheology on a revolving CD based device. Most important aspect of this investigation is to bring out the role of the fluid rheology towards dictating the critical Reynolds number and rotation number for the flow. Apart from that, we also attempt to bring out the structures of the secondary vortices mediated by the underlying instabilities, bearing potential key implications in realizing enhanced micromixing and rapid diagnostics of diseases on a portable rotational platform.

The article is organized as follows: In section 2, the governing equations and linear stability formulation of the shear-thinning flow in a rotating platform are presented. The stability characteristics are addressed in section 3, by considering a shear-thinning fluid subjected to infinitesimal spanwise disturbances and homogeneous streamwise disturbances. The analysis is focused on the temporal growth of spanwise perturbations, marginal stability boundaries and structure of roll-cell formed, with and without viscosity perturbation. Neutral stability boundaries for different shear-thinning indices and time constants are drawn to characterize transition to instabilities. Finally, conclusions are summarized in section 4.

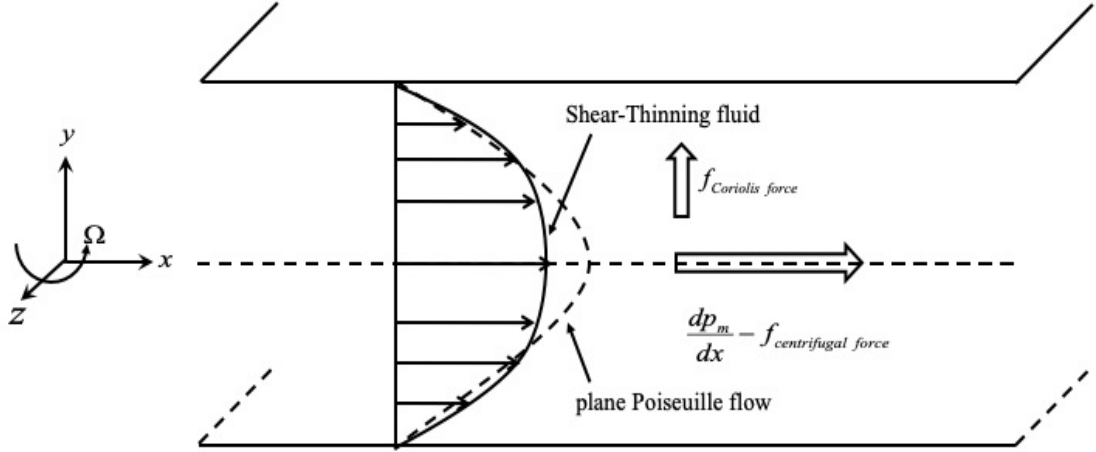

Figure 1. Schematic of the three-dimensional non-Newtonian fluid through a microchannel, rotating with a constant angular velocity ($\Omega$) about the z-axis which is perpendicular to the plane of the paper.

## 2. Mathematical modeling

### 2.1 Governing equations

We consider the flow of an incompressible shear-thinning fluid between two parallel plates at $y = \pm D_h$, rotating at a constant speed ($\vec{\Omega} = \Omega \hat{k}$) on a revolving CD based device, about the z-axis (see figure 1). The flow is actuated due to centrifugal force acting on the fluid, thus no external source; for example, pump is not required to trigger the flow. We use the half channel width ($D_h$), the steady-state maximum velocity ($U_{max}$) to non-dimensionalize the governing equations for the system under investigation. The fluid pressure is modified by incorporating the centrifugal force and thus termed as modified pressure ($p_m$)[20,54]. The modified pressure is scaled using $\rho U_{max}^2$, where $\rho$ is the fluid density. Using the above-mentioned scales, the non-dimensional form of the governing equation reads as:

$$\nabla \cdot \vec{u} = 0 \qquad (1.1)$$

$$\frac{\partial \vec{u}}{\partial t} + (\vec{u} \cdot \nabla)\vec{u} = -\nabla p_m + \nabla \cdot \vec{\tau} - \vec{F}_c \qquad (1.2)$$

where $\vec{u} = (u, v, w)$ denotes the fluid velocity, $\vec{F}_c$ is Coriolis force given by $\vec{F}_c = Ro(-v, u, 0)$ with $Ro$ is the rotation number defined as $Ro = \Omega D_h / U_{max}$ and $\vec{\tau}$ denotes the deviatoric stress tensor. Here, fluid apparent viscosity depends on the shear rate and the constitutive relationship illustrates as follows:

$$\bar{\tau} = \frac{1}{\overline{Re}} \mu(\Gamma) \dot{\gamma} \qquad (1.3)$$

where,

$$\dot{\gamma} = \nabla \vec{u} + (\nabla \vec{u})^T \qquad (1.4)$$

and the Reynolds number for the considered flow is $\overline{Re}$, defined as $\overline{Re} = \rho U_{max} D_h / \mu_{avg}$. It may be noted that to compare our result with the reported literature, we have defined the Reynolds number based on average viscosity ($\mu_{avg}$) across the channel, which is logical choice for viscosity-stratified flows. Such a definition Reynolds number was suggested by Wall and Wilson (1996)[55] for Newtonian fluids and later adopted by Chikkadi (2005)[56] and Nouar (2007)[57] for Carreau fluids. Further, in Eq. (1.3), the variable $\dot{\gamma}$ is the rate-of-strain tensor and $\Gamma$ its second invariant: $\Gamma = \frac{1}{2}\sqrt{(\dot{\gamma} : \dot{\gamma})}$ (: denotes the dyadic product of two tensors). Here, the viscosity of the shear-thinning fluid is modelled using the Carreau model proposed by Carreau (1972),[44]

$$\frac{\mu - \mu_\infty}{\mu_0 - \mu_\infty} = \left(1 + (\lambda \Gamma)^2\right)^{\frac{n-1}{2}}, \qquad (1.5)$$

where $\mu_0$ and $\mu_\infty$ are the viscosity of the proposed fluid at low and high shear rate and $n$ is the shear-thinning index $(n<1)$, and $\lambda$ is the non-dimensional characteristic time constant of the fluid ($D_h/U_{max}$ is used for non-dimensionalize $\lambda$). It is worth mentioning that the viscosity at high shear rate is significantly small and thereby the ratio $\mu_\infty/\mu_0$ is negligible, hence neglected for all future calculations [58,59]. Thus, the dimensionless form of effective viscosity is given by:

$$\frac{\mu}{\mu_0} = \left(1 + (\lambda \Gamma)^2\right)^{\frac{n-1}{2}}. \qquad (1.6)$$

Johnston et al. (2004)[51] showed that at low shear rate, generalized power law,[37] Power law and Carreau models[35,60] effectively converge to the strain rate dependent apparent viscosity behavior as depicted by Eq. (1.6).

**2.2 Base flow**

We are interested in exploring the linear instability of the steady unidirectional base flow $\vec{U}_b = (U_b, 0, 0)$ satisfying the system of equations (1.1)-(1.2) with no-slip condition at the rigid walls. The non-dimensional base velocity profile can be obtained from the $x$-momentum equation as follows : [33,56]

$$-\overline{Re}\frac{dp_m}{dx} + \frac{d}{dy}\left(\mu \frac{dU_b}{dy}\right) = 0. \qquad (1.7)$$

The base flow equation 1.7 is solved numerically using built-in routines bvp4c of MATLAB and then subsequently transformed into Chebyshev space using a spline interpolation formula. We have compared our computed base flow profiles with the results of Nouar (2007) for time constant $\lambda = 10.0$ and three different shear-thinning indices $n = 0.3, 0.7$ and $1.0$ in figure 2(a).

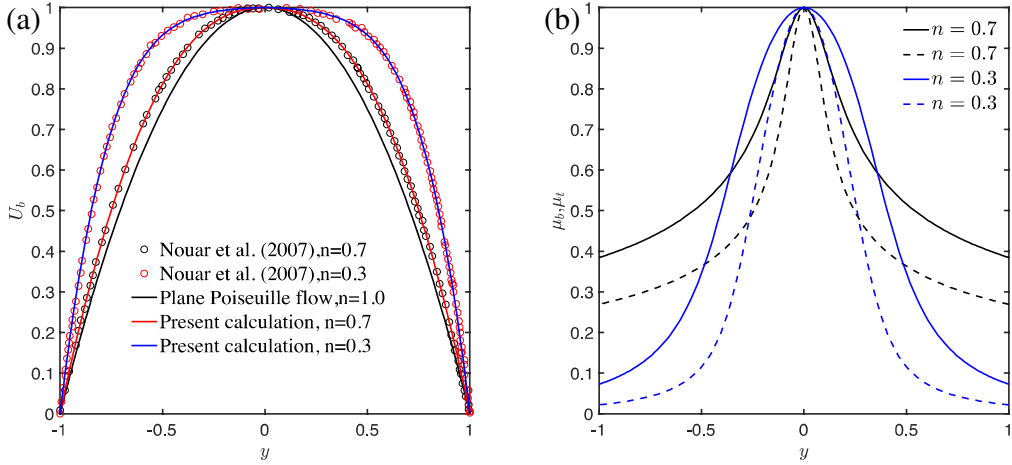

Figure 2. (a) Base velocity profiles for different shear-thinning index $n = 0.3, 0.7, 1.0$. The 'o' sign corresponds to the results of Nouar (2007). (b) Solid line for the base viscosity profile $(\mu)$ and dashed lines represent the tangent viscosity $(\mu_t)$. The used time constant is $\lambda = 10.0$.

In figure 2(b), we also have shown the variation of $\mu$ and $\mu_t$ (tangent viscosity) throughout the channel, where $\mu_t$ is defined as the rate of change of shear-stress with respect to shear strain $(d\tau_{xy}/d\dot{\gamma}_{xy})$ of the mean flow direction.[32,33] Our computed base profiles are in good agreement with the results of Nouar (2007). It is notable to mention that the Carreau model behaves like Newtonian fluid with constant viscosity for $n = 1$ or $\lambda = 0$.

### 2.3 Linear stability equations

To understand the effects of the infinitesimal disturbance $(\vec{u}', p')$ on the base flow, the momentum equations are linearized around the base flow and modified pressure, which leads to:

$$\frac{\partial \vec{u}'}{\partial t} = -(\vec{u}'.\nabla)\vec{U}_b - (\nabla.\vec{U}_b)\vec{u}' - \nabla p' + \frac{1}{\overline{Re}}\nabla\overline{\overline{\tau}}' - Ro\vec{F}_c' \qquad (1.8)$$

where, the shear stress of perturbations $(\vec{\tau}')$ is given by $\vec{\tau}' = \mu(\vec{U}_b)\dot{\gamma}(\vec{u}') + \mu'\dot{\gamma}(\vec{U}_b)$, and $\vec{F}_c' = (-v', u', 0)$. The viscosity perturbation,[32,33] $\mu'$ is denoted by :

$$\mu' = \dot{\gamma}_{ij}(\vec{u}')\frac{\partial \mu}{\partial \dot{\gamma}_{ij}}(\vec{U}_b). \qquad (1.9)$$

Since we have considered the base flow to be unidirectional, it can be shown that

$$\tau'_{ij} = \mu(\vec{U}_b)\dot{\gamma}_{ij}(\vec{u}') \text{ for } ij \neq xy, yx, \qquad (1.10)$$

$$\tau'_{ij} = \mu_t(\vec{U}_b)\dot{\gamma}_{ij}(\vec{u}') \text{ for } ij = xy, yx, \qquad (1.11)$$

where,

$$\mu_t = \mu(\vec{U}_b) + \frac{d\mu}{d\dot{\gamma}_{xy}}(\vec{U}_b)\dot{\gamma}_{xy}(\vec{U}_b) \qquad (1.12)$$

represents the tangent viscosity. For shear-thinning fluids, it is observed that $\mu_t < \mu$ which can affect the flow stability property immensely. The amplitude-wave harmonic form of the three-dimensional is assumed as $[\vec{u}', p'] = [\vec{\tilde{u}}(y), \tilde{p}(y)]\exp(i(\alpha x + \beta z - \omega t))$ and $\vec{\tilde{u}}(y) = (\tilde{u}, \tilde{v}, \tilde{w})$ where, $\alpha$ and $\beta$ are the streamwise and spanwise wave numbers and $\omega$ is the complex temporal frequency.[61] The equation (1.8) can be written in terms of normal velocity $\tilde{v}$ and vorticity $\tilde{\eta} = i\beta\tilde{u} - i\alpha\tilde{w}$, as follows:

$$-i\omega\begin{pmatrix}\Delta\tilde{v}\\ \tilde{\eta}\end{pmatrix} = \begin{pmatrix}L_{os} & \ell_{12}\\ \ell_{21} & L_{sq}\end{pmatrix}\begin{pmatrix}\tilde{v}\\ \tilde{\eta}\end{pmatrix}, \qquad (1.13)$$

where, the operators $L_{os}$, $\ell_{12}$, $\ell_{21}$ and $L_{sq}$ are defined as

$$L_{os} = \alpha\left(U\Delta - D^2 U\right) + \frac{i}{Re}\left(\mu\Delta^2 + 2D\mu D^3 + D^2\mu D^2 - 2k^2 D\mu D + k^2 D^2\mu\right)$$
$$+ i\underbrace{\frac{\alpha^2}{Rek^2}\left(D^2 + k^2\right)\left[(\mu_t - \mu)\left(D^2 + k^2\right)\right]}_{term\,I}, \qquad (1.14)$$

$$\ell_{12} = -i\underbrace{\frac{\alpha\beta}{Rek^2}\left(D^2 + k^2\right)\left[(\mu_t - \mu)D\right]}_{term\,IV} - iRo\beta, \qquad (1.15)$$

$$\ell_{21} = \underbrace{-i\beta\left(DU - Ro\right)}_{term\,III} - i\frac{\alpha\beta}{Rek^2}D\left[(\mu_t - \mu)\left(D^2 + k^2\right)\right], \qquad (1.16)$$

$$L_{sq} = \alpha U + \frac{i}{Re}\mu\Delta + \frac{i}{Re}D\mu D + \underbrace{\frac{i}{Re}\frac{\beta^2}{k^2}D[(\mu_t - \mu)D]}_{\text{term II}}. \qquad (1.17)$$

with $k^2 = \alpha^2 + \beta^2$, $D = d/dy$ and $\Delta = D^2 - k^2$.

As per literature, no exact direct equivalence of Squire theorem is available for a non-Newtonian fluid with viscosity perturbation. However, several tests have been performed by earlier researchers which indicate that the lowest critical Reynolds number can be obtained for streamwise disturbances (i.e. for $\beta = 0.0$). Keeping this knowledge in mind, one can infer that, by help of suitable scaling, the Squire theorem may applicable for the non-Newtonian fluid with viscosity perturbation.[33,57] Else ways, without viscosity perturbation, it can be easily shown that the Squire theorem is valid and an equivalent two-dimensional problem can be defined.[61]

**2.4 Numerical Methodology**

In order to solve the coupled Orr-Sommerfeld and Squire equations boundary value problem described in the equation (1.13) with the boundary condition $\tilde{v}(\pm 1) = D\tilde{v}(\pm 1) = \tilde{\eta}(\pm 1) = 0$, we have followed and implemented the methodology as described in Trefethen (2000).[62] The flow domain is discretized using Chebyshev grids which eventually suggests to consider the vectors of unknowns to specific the normal velocity and vorticity $(\tilde{v}, \tilde{\eta})$ of the equation (1.13) at the grid points in the following manner: $(\tilde{v}_0, \tilde{v}_1, \tilde{v}_2, ..., \tilde{v}_N, \tilde{\eta}_0, \tilde{\eta}_1, \tilde{\eta}_2, ..., \tilde{\eta}_N) = X$ (say), where subscripts 0 and $N$ denote the boundary points. In most of the hydrodynamic stability problems, the node sets satisfying the density property (means clustering near the boundary points) on the bounded interval $[-1,1]$ are called Chebyshev points. There exist different types of collocation points that can be applied to discretize a flow domain depending on the boundary condition and complexity of the problem. Out of them, the most commonly used one is Chebyshev Gauss-Lobatto quadrature points:

$$y_j = \cos\left(\frac{j\pi}{N}\right), \quad j = 0, 1, 2, ..., N. \qquad (1.18)$$

The space derivative present in the eigenvalue problem (1.13) needs to be approximated using numerical differentiation techniques. The higher order differentiation matrix can be constructed from the first order differentiation matrix by making use of the fact that $D^{(m)} = (D^{(1)})^m$ for $m = 2, 3, ...$. Finally, using the differentiation matrix of required order in the equation (1.13), obtained a system of linear equation, which corresponds an eigenvalue problem of the form $AX = -i\omega BX$. The coefficient matrices $A$ and $B$ are complicated $2(N+1) \times 2(N+1)$ order matrices with four different blocks. To implement the clamped boundary conditions, the rows in the block matrices corresponding to the boundary points of normal velocity and vorticity vectors are removed, and thus final order of the coefficient matrices $A$ and $B$ gets reduced to $2(N-1)\times 2(N-1)$. The eigenvalues of the coefficient matrices are found using built-in subroutine in the MATLAB and for each set of character parameters $(n,\lambda)$, the Reynolds number ($\overline{Re}$), rotation number ($Ro$) and wave numbers ($\alpha,\beta$) are varied such that $|\omega_i|\leq 10^{-6}$.

## 3. Results and Discussions

### 3.1 Numerical Validation

The accuracy of the results obtained from our developed code against Chikkadi (2005) and Nouar (2009) is tested in figure 3(a), where critical Reynolds number, $\overline{Re}_{cr}$ is reported as a function of time constant $\lambda$ for a shear-thinning index $n=0.5$. Results are compared for both with ($\mu'\neq 0$) and without ($\mu'=0$) viscosity perturbation. Our numerical results show an excellent agreement with Chikkadi (2005) as well as Nouar (2009). We also compared our results with reported literature on the Newtonian flows with rotating body force.[63] The Carreau law model describes the behavior of Newtonian fluid under the limits: (i) $n=1$ and $\lambda\neq 0$ in the figure 3(b), and (ii) $n\neq 1$ and $\lambda=0$ in figure 3(c). The Reynolds number is fixed at $\overline{Re}=10,800$ (for the case of Newtonian fluid, the average viscosity is constant and the definition of Reynolds number remains unaffected with the present scaling).

In Table-$I$, we have listed the rheological parameters used for Carreau model reported in the literature to assess the behavior of blood flow accurately. It is to note that the rheological parameters in table-$I$ is validated by Banerjee et al. (2008) from the experimental datas of blood analog fluid.[64] The same sets of parameters is utilized to estimate the steady unidirectional base flow, viscosity profile etc., in order to examine the rotational effect on the onset of instability of blood flow through a microchannel mounted on a revolving CD based device.

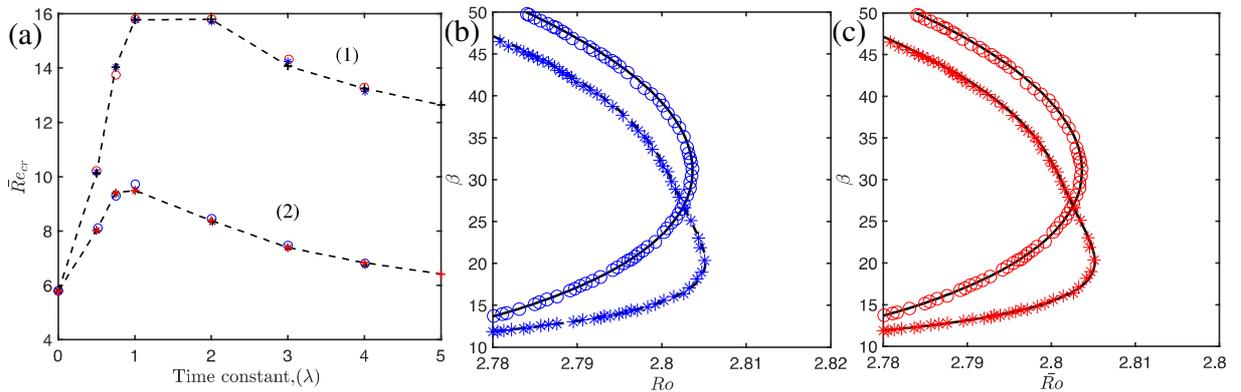

Figure 3. (a) Critical Reynolds number as a function of time constant ($\lambda$) for shear-thinning index $n=0.5$: line (1) corresponds to excluding viscosity perturbation, and line (2) corresponds

to including viscosity perturbation ('o' and '*' sign corresponding to Nouar (2007) and Chikkadi (2005), respectively, and '+' is for present computation). Neutral curves for rotating flow with Newtonian limit: (b) $n = 1, \lambda \neq 0$ and (c) $n \neq 1, \lambda = 0$ at Reynolds number $10,800$ ('o' and '*' sign representing the results of Wallin (2013) for streamwise wave number $\alpha = 0$, $\alpha = 2.7$, respectively, and the solid lines display present computation results).

Samples for the base velocity profile, velocity gradient and base viscosity and tangent viscosity profile are shown in figure 4 for four different values of the time constant $\lambda = 0.5, 0.75, 1.5, 2.0$ and two different shear-thinning index $(n = 0.2, 0.3568)$. At lower range of for two different shear-thinning index $(n = 0.2\,(\text{blue line}),\, 0.3568\,(\text{black line}))$. The values of time constant varies as $\lambda = 0.5$, $\lambda = 0.75$, $\lambda = 1.5$ and $\lambda = 2.0$ from left to right.

| Literature | $n$ (---) | $\lambda$ ($s$) | $\mu_0$ (cP) | $\mu_\infty$ (cP) | $\dfrac{\mu_\infty}{\mu_0}$ |
|---|---|---|---|---|---|
| Banerjee et al., (2008) | 0.2 | 9.56 | 55.0 | 3.39 | 0.061636 |
| Cho and Kensey, (1991) | 0.3568 | 3.313 | 56.0 | 3.6 | 0.0625 |

Table-I. The parametric values for non-Newtonian constitutive relationships for blood flow.

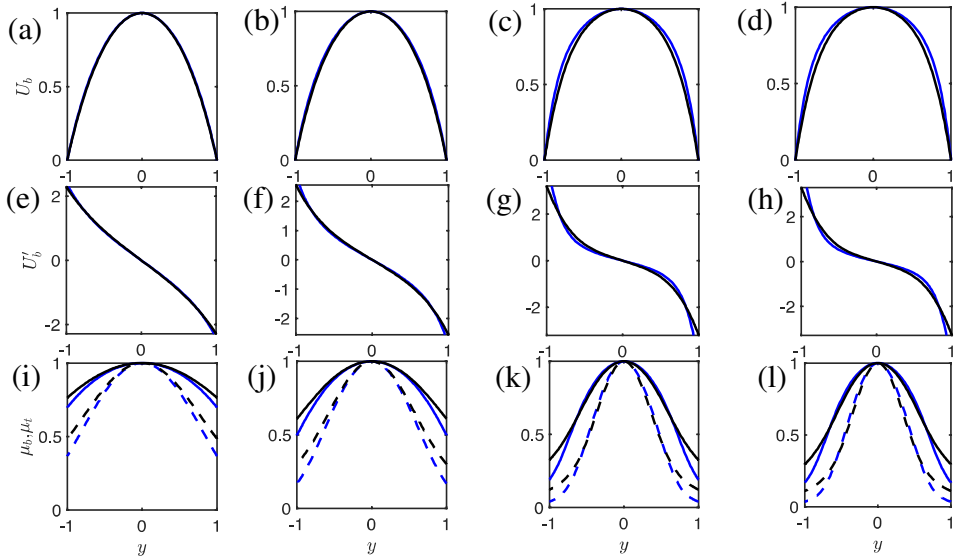

Figure 4. Steady unidirectional base velocity profiles in (a)-(d), velocity gradient curves in (e)-(h), and in (i)-(l) the viscosity (solid line) and tangent viscosity (dashed line) profiles are shown

time constant $(\lambda = 0.5)$, the shear-thinning index $n$ has minor influence on the base velocity and its gradient. However, the base velocity profile becomes 'fuller' (see figure 4(c) and (d) with the values of time constant as $\lambda = 1.5$ and $\lambda = 2.0$, respectively) for increasing shear-thinning index. However, for a lower range of the time constant $(\lambda)$, the upshot is not so prominent. It is worth noting that, a smaller $n$ and higher $\lambda$ can affect to increase the shear thinning behavior of non-Newtonian fluid. Gradients of base velocity curves in figure 4(e)-(h) suggest, unlike plane Poiseuille flow where the velocity gradient variation is linear, for shear-thinning fluid like blood, the variation is highly non-linear because of a strong coupling between velocity and viscosity. The scenario depends on time constant as well as shear-thinning indices. This non-linear variation also alters the stability characteristics of Coriolis force-driven instability. As observed from the figure 4(i)-(l), the base viscosity is inferior for lower value of $n$ and continuously decreases towards the wall.

In the Table-*II* below, the average viscosity $(\mu_{avg})$, tangent viscosity $(\mu_{t,avg})$ and $(\mu_{avg} - \mu_{t,avg})$ are listed for different shear-thinning indices $(n)$ and time constants $(\lambda)$. It can be observed that the difference $(\mu_{avg} - \mu_{t,avg})$ is monotonically increasing with respect to the time constant, when shear-thinning index $n$ is fixed. This is mainly because of the fact that for shear-thinning fluids, contribution of the term $d\mu/d\dot{\gamma}_{xy}(\vec{U}_b)\dot{\gamma}_{xy}(\vec{U}_b)$ is negative for tangent viscosity.

| $n$ | $\lambda$ | $\mu_{avg}$ | $\mu_{t,avg}$ | $\mu_{avg} - \mu_{t,avg}$ |
|---|---|---|---|---|
| 0.2 | 0.5 | 0.901497468686692 | 0.760030550228618 | 0.141466918458074 |
| 0.2 | 0.75 | 0.834863456509516 | 0.649296981891198 | 0.185566474618318 |
| 0.2 | 1.5 | 0.717447059850853 | 0.514955403604216 | 0.202491656246636 |
| 0.2 | 2.0 | 0.675260656808234 | 0.476455516832211 | 0.198805139976023 |
| 0.3568 | 0.5 | 0.919594762397429 | 0.800766963326587 | 0.118827799070842 |
| 0.3568 | 0.75 | 0.861484413241518 | 0.697176780082068 | 0.164307633159450 |
| 0.3568 | 1.5 | 0.740218029197085 | 0.541712904639920 | 0.198505124557165 |
| 0.3568 | 2.0 | 0.689891785635261 | 0.490620403472551 | 0.199271382162709 |

Table-*II*. Average viscosity, tangent viscosity and $(\mu_{avg} - \mu_{t,avg})$ for two different shear-thinning index (*n*) and time constant $\lambda = 0.5, 0.75, 1.5$ and $2.0$.

### 3.2 Stability curves and critical parameters

Neutral stability curves in $\beta - \overline{Re}$ plane are shown in figure 5(a)-(d) for two different shear-thinning indices: $n = 0.2$, $0.3568$ and time constants $\lambda = 0.5$. The results in left and right columns of the figure 5 are, respectively, for the flow configuration including viscosity perturbation ($\mu' \neq 0$) and excluding viscosity perturbation ($\mu' = 0$). An increase in the rotation number destabilizes the flow by lowering the Reynolds number $(\overline{Re})$. The smallest critical Reynolds number, with or without viscosity perturbation, occurs for $Ro = 0.34$ and $0.33$, respectively. However, increasing the rotation number further, a stabilizing influence is observed by means of higher critical Reynolds number in $\beta - \overline{Re}$ plane. The rotation number hallmarking the onset of such a phenomenon is designated as the critical rotation number $(Ro_{cr})$. The values of critical rotation number, for the set of rheological parameters $(n, \lambda) = (0.2, 0.5)$ and $(0.3568, 0.5)$, are found to be $0.34$ and $0.33$, respectively. It is worthwhile to mention that the value of $Ro_{cr}$ is not constant and changes with the different values of the rheological parameters $(n, \lambda)$. Further discussion on this point is made later on in the section below.

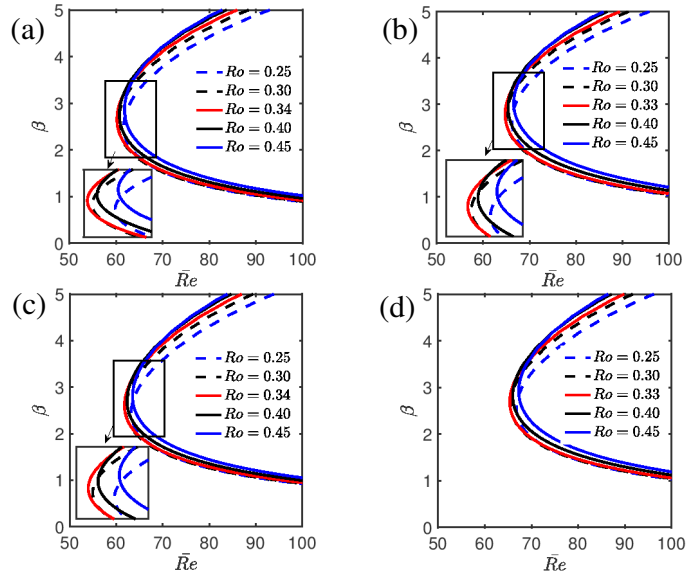

Figure 5. Stability boundaries for two different shear-thinning indexes when $\lambda = 0.5$: in (a), (b) $n = 0.2$ and in (c), (d) $n = 0.3568$. Left and right columns are with and without viscosity

perturbation, respectively. Solid red line is used to indicate critical rotation number for the above flow configuration with two-dimensional homogeneous streamwise disturbances $(\alpha = 0)$.

Viscosity stratification as a possible means of stabilizing the flow is well known in the reported literature.[56,57,65,66] The critical Reynolds number obtained from neutral stability curves with viscosity perturbation, as shown in the figures 6(a) and (c), is found to be lower than that of without viscosity perturbation (see figure 6(b) and (d)). Interestingly, our computation confirms that the $Ro_{cr}$ remains almost unaltered for the flow with and without viscosity perturbation (for the set of parameter range $(n, \lambda)$ considered herein). This confirms that the destabilization mechanism associated with the Coriolis force based instability is independent of viscosity perturbation for zero streamwise wavenumber $(\alpha = 0)$. Moreover, this mechanism is reflected mainly by the *term III* in equation (1.16) and not by *term IV* in equation (1.15). As per earlier investigations on the shear-thinning flow through a non-rotating channel, the critical Reynolds number in case of two-dimensional disturbances $(\alpha \neq 0, \beta = 0)$ with viscosity perturbation is found to be half than that of the flow without viscosity perturbation.[56,57] This is because of the fact that the viscosity perturbation produces a positive definite term (*term I* as marked in (1.14)), which in turn reduces the viscous dissipation compared to the flows without viscosity perturbation.[57] In effect, this delays the onset of instability when an infinitesimal two-dimensional streamwise disturbance is introduced with the base flow. However, the numerical computation confirms that the above degree of destabilization in terms of critical Reynolds number is unachievable for two-dimensional spanwise disturbances. The magnitude of critical Reynolds number obtained for spanwise system of rotation $(\alpha = 0)$ in $\beta - \overline{Re}$ plane with and without viscosity perturbation is of similar order. The reason of this discrepancy is that for two-dimensional streamwise perturbation, Tollmien–Schlichting (TS) mode is mainly responsible for transition to occur. However, for a spanwise two-dimensional disturbance, in a rotating channel the destabilization mechanism is strongly dependent upon rotational force and velocity gradient of base flow. The higher growth rate due to spanwise disturbances reduces the magnitude of critical Reynolds number considerably.[11,67,68]

In figure 6(a)-(d), the neutral stability boundaries are shown by varying the time constant $\lambda$, with (see figure 6(a) and (c)) and without (see figure 6(b) and (d)) viscosity perturbation, for two different shear-thinning indices $n = 0.2$, $0.3586$. The stability curves are obtained herein at the respective critical rotation numbers for each pair of rheological parameters $(n, \lambda)$. The critical rotation numbers for shear-thinning index $n = 0.2$ and time constant $\lambda = 0.5, 0.75, 1.0, 1.25, 1.5, 1.75$ and $2.0$ are $Ro_{cr} = 0.34, .33, 0.21, 0.29, 0.27, 0.24$ and $0.22$ respectively, and for $n = 0.3568$ are $Ro_{cr} = 0.34, 0.33, 0.31, 0.29, 0.27, 0.24$ and $0.22$ for the case of $\mu' \neq 0$. The effect of shear-thinning (increasing $\lambda$) for spanwise system of rotation is progressively destabilizing in the $\beta - \overline{Re}$ plane for shear-thinning index $n = 0.2$ (see figure 6(a) and (b)), which is in contrast to earlier findings for streamwise two-dimensional disturbances.[32,56] Although the critical Reynolds number is not distinguishable in the figure 6(c) and (d), the same can be concluded for shear-thinning index $n = 0.3568$. Increasing value of $\lambda$ decreases the critical rotation number, which implies that lesser effort is required to overcome the stabilizing effect of viscosity and destabilize the flow. The critical Reynolds number for $\mu' \neq 0$ is less than that of the corresponding $\mu' = 0$ case. This is because of the fact that the *term II* in the equation (1.17) is positive definite which gives a reduction in the viscous dissipation as compared to the consideration excluding viscosity perturbation. Consequently, the onset of instability is advanced for the two-dimensional spanwise rotating channel flow by introducing infinitesimal perturbation into the flow system. Similar kind of result has also been obtained by earlier researchers for two-dimensional streamwise perturbation.[32]

Next, we intend to explore the effect of Coriolis force based instability mechanism on the shear-thinning microchannel flow and thereby assess the factors like the non-linear viscosity of shear-thinning fluid, base velocity profile, velocity gradient, curvature of base velocity, viscosity perturbation that influence the stability characteristics. In figure 7, we have shown the neutral stability boundaries for four different values of $\lambda$ $(0.5, 0.75, 1.5$ and $2.0)$, with spanwise two-dimensional disturbances $(\alpha = 0)$, in the $\beta - \overline{Re}$ plane to understand the outcome of the variation of shear-thinning indices $n$ for two different cases (i) $\mu' \neq 0$ (see figure 7(a)-(d)) and (ii) $\mu' = 0$ (see figure 7(e)-(h)). Note that, here we are exploring the instability behavior due to Coriolis based instability mechanism for two-dimensional disturbance $(\alpha = 0)$. Hence, the effect of base flow, second derivative of viscosity and the terms originating due to viscosity perturbation appearing in the modified Orr-Sommerfeld and Squire operator ((1.14)-(1.17)) have no contribution in the instability mechanism (as $\alpha = 0$) for both the case of $\mu' = 0$ and $\mu' \neq 0$.

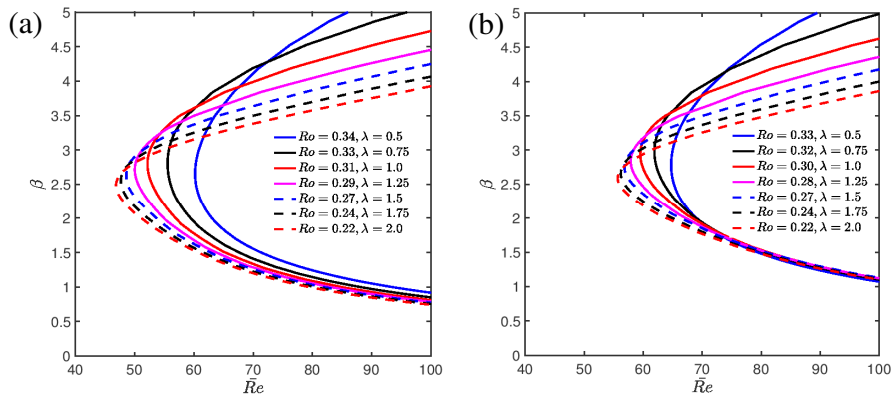

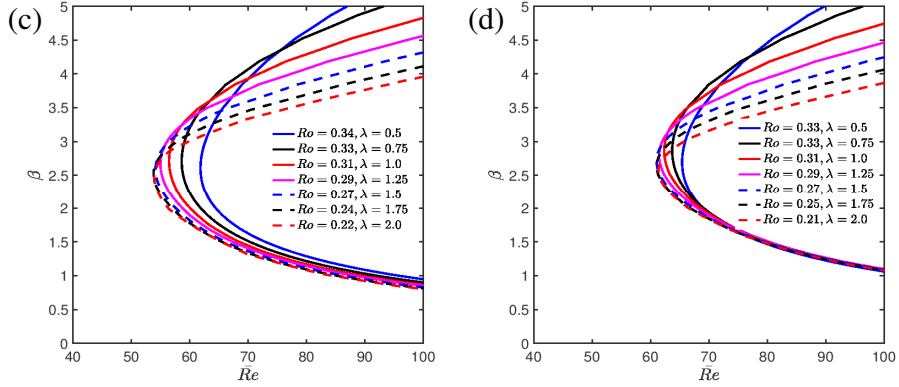

Figure 6. Neutral stability curves: (a) without viscosity perturbation, (b) with viscosity perturbation for shear-thinning index $n = 0.2$, and (c) without viscosity perturbation, (d) with viscosity perturbation for $n = 0.3568$.

The only term that alters the neutral stability boundary for two different shear-thinning indices considered herein is the velocity gradient (*term III* of equation (1.16)) and viscosity distribution of the fluid. This shows that increasing shear-thinning index makes the flow slightly more stable, and hence a delay in the onset of transition for this flow.

In the figure 8, we have calculated the critical parameters, namely $\beta_{cr}, Re_{cr}$ as a function of the time constant ($\lambda$) under consideration, for both $\mu' \neq 0$ and $\mu' = 0$. Independent of the shear thinning index ($n$), the shorter waves are found to be critical as the time constant increases. The critical value of the wave number is almost around the time constant 0.75 and thereafter tends to decrease. Thus, the 'shorter' wavelength is most critical for the shear-thinning viscosity at $\lambda = 0.75$ (figure 8(a)). Figure 8(b) shows variation of the critical Reynolds number as a function of the time constant $\lambda$. Although the critical value of Reynolds number is not changing drastically, the monotonically decreasing behavior of the curves suggests a destabilizing effect of increasing $\lambda$ on the onset of instability mechanism, and it is due to the decrease in average viscosity (averaged over wall-normal direction) of the system for higher time constant.

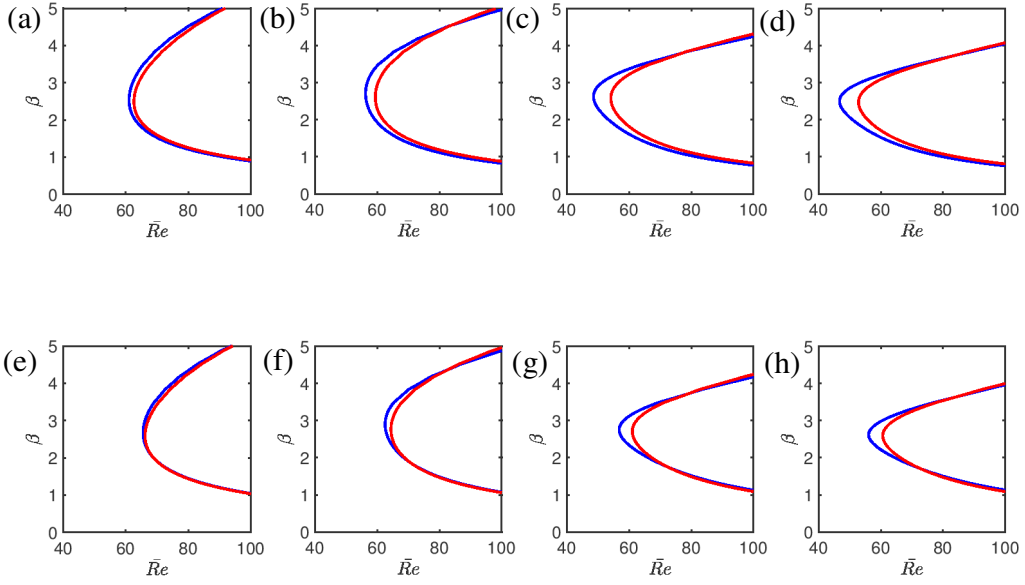

Figure 7. Neutral stability curves for $\mu' \neq 0$ with (a) $\lambda = 0.5$, (b) $\lambda = 0.75$, (c) $\lambda = 1.5$ and (d) $\lambda = 2.0$, and for $\mu' = 0$ with (e) $\lambda = 0.5$, (f) $\lambda = 0.75$, (g) $\lambda = 1.5$ and (h) $\lambda = 2.0$. All the stability curves are drown for a rotation number $Ro = 0.30$.

### 3.3 Structure of the roll-cells

One of the predominant characteristics of the rotating flow system is the generation of secondary flow as influenced by passive means like Coriolis force, which enhances fluid mixing, heat exchange etc., and facilitates applications such as trapping and seperation of cells or particles.[69,70] The flow velocity in the axial direction is determined by the centrifugal force along the channel. Moreover, the secondary flow is mainly regulated due to Coriolis force. We are interested to understand the structure of roll-cell generated due to secondary flow transition in the $y-z$ plane owing to the spanwise system of rotation and streamwise homogenious disturbances $(\alpha = 0)$. Different values of time constants, $\lambda = 0.50, 0.75, 1.50, 2.0$ are considered by keeping the shear-thinning index $n = 0.20$. We typically choose the shear-thinning index $n = 0.20$ because the early onset of transition is observed for the present choice. Hence, it is interesting to study the structure of the roll-cells forming due to secondary flow generated by implementing spanwise disturbance. Figures 9(a)-(d) correspond to the roll-cell patterns and streamlines for the critical parameters obtained for shear-thinning fluid considered herein. In all the cases, the roll-cells are characterized by two distinct vortices rotating in opposite directions. The centers of the vortices are near to the leading edge (bottom wall), which is the destabilizing side due to an unstable "stratification" because of the Coriolis force.[68] Depending on viscous and Coriolis force, the cellular patterns formed at $\lambda = 0.50$ (see figure 9(a)) are mostly confined towards the leading edge (bottom wall) of the channel. It is quite interesting that due to effect of the Coriolis force, another maxima in the longitudinal speed of the streak occurs slightly lower than the central line of the channel. At lower value of the time constant, the vortices are forced to be confined within the bottom region. As the shear-thinning effect increases, the critical Reynolds number for the onset to transition decreases and lesser rotational effort is required to destabilize the flow. The corresponding roll-cell patterns are no longer confined to the bottom region and seem to be occupying the whole fluid layer. Moreover, the cellular patterns are supposed to be stretched from leading edge to trailing edge as the shear-thinning effect increases. This information in effect can potentially be used for enhancing in mixing and/or separation process depending on the specific application on hand.

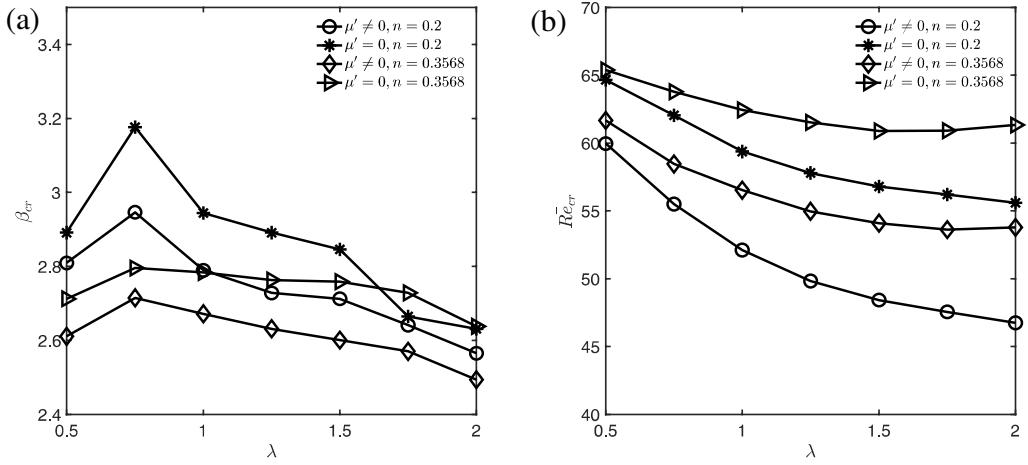

Figure 8. (a) critical spanwise wave number and (b) critical Reynolds number as a function of time constant ($\lambda$) for two different shear thinning indices ($n$). Results are drown for both with and without viscosity perturbation. All results are obtained for the streamwise wave number $\alpha = 0.0$ and corresponding $Ro_{cr}$.

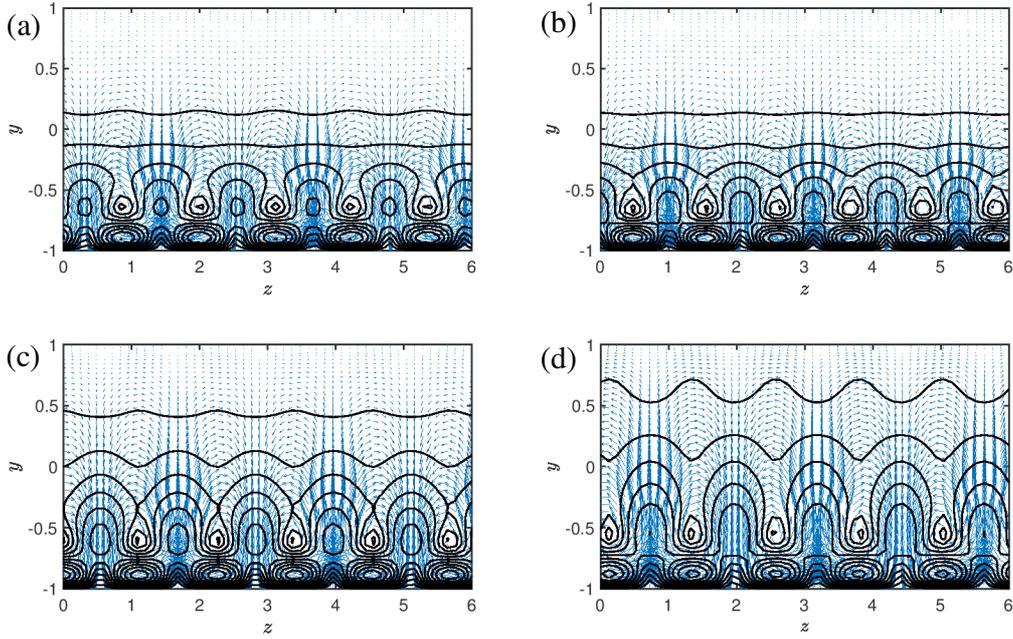

Figure 9. Structure of the roll-cell for shear-thinning index $n = 0.2$ and four different time constant at respective critical parameters (a) $\lambda = 0.50$, (b) $\lambda = 0.75$, (c) $\lambda = 1.5$ and (d) $\lambda = 2.0$. The other critical parameters are $(2.81, 59.96, 0.34)$, $(2.95, 55.51, 0.33)$, $(2.71, 48.43, 0.27)$ and $(2.56, 46.75, 0.22)$ respectively in the following order $(\beta, \overline{Re}, Ro)$.

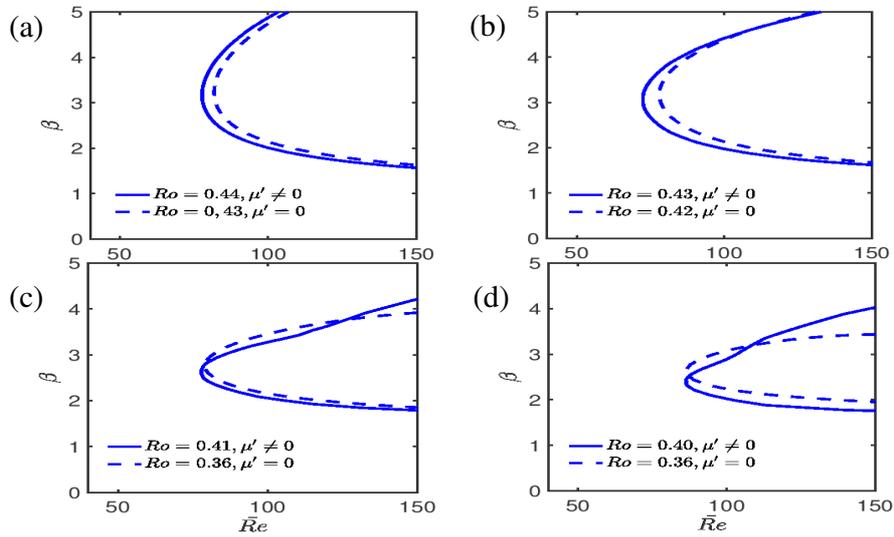

Figure 10. Neutral stability courves in $\beta - \overline{Re}$ plane corresponding to three dimensional disturbances with streamwise wave number $\alpha = 1.0$ for shear-thinning index $n = 0.20$ and four different time constant (a) $\lambda = 0.5$, (b) $\lambda = 0.75$, (c) $\lambda = 1.50$ and (d) $\lambda = 2.0$ at the respective critical rotation number.

### 3.4 Effect of three-dimensional disturbances

Our discussion thus far provides two-dimensional (2D) instability behaviors with stream wise wave number $\alpha = 0.0$ and does not give any conclusion about the three-dimensional (3D) instability of the considered flow. In Figure 10, the sets of neutral stability curves are plotted in the $\beta - \overline{Re}$ plane for 3D perturbations (with both $\mu' \neq 0$ and $\mu' = 0$), for shear-thinning index $n = 0.20$ and stream wise wave number $\alpha = 1.0$. The results are plotted for both the cases $\mu' \neq 0$ and $\mu' = 0$ at the respective critical rotation numbers. Compared to 2D disturbances, much higher rotational effort is required for the onset of instability to occur in the case of 3D disturbances. In order to understand why such a scenario is happening, we multiplied the Squire equation by the complex conjugate of the eigenfunction corresponding to the most unstable eigenmode of the operators (1.16) and (1.17) to arrive the following (by taking the real part of the both side):

$$\omega_i <|\tilde{\eta}|^2> = \underbrace{-\frac{1}{\overline{Re}} < \mu(|D\tilde{\eta}|^2 + k^2|\tilde{\eta}|^2)>}_{term\,V} + \underbrace{\frac{\beta^2}{k^2 \overline{Re}} < (\mu - \mu_t)|D\tilde{\eta}|^2 >}_{term\,VI} - \beta \underbrace{< (U' - Ro)|\tilde{v}|^2 >}_{term\,VIII}$$
$$+ \underbrace{\frac{\alpha\beta}{\overline{Re}} \{< (\mu - \mu_t)(|D\tilde{v}|^2 - k^2|\tilde{v}|^2)> + <(\mu - \mu_t)(\tilde{v}_r D\tilde{v}_r + \tilde{v}_i D\tilde{v}_i)>\}}_{term\,VII},$$
(1.19)

where, $<\phi>$ represents the integral $\int_{-1}^{1} \phi \, dy$. It is clear from (1.19) that the perturbed flow becomes unstable if the contribution from the $term\,VIII$ mentioned as above becomes negative and higher than the $term\,V$. This reduces to the criteria set by Bradshaw (1969) and independently by Pedley (1969), in the specific case of parallel shear flows subject to rotation. In the case of 3D perturbations, the $term\,V$ contributes more, owing to the fact that $k^2 = \alpha^2 + \beta^2$. In addition, for non-zero streamwise wave number, the positive definite contribution from the viscosity is higher. In the case of $\mu' \neq 0$, for shear-thinning fluid $\mu > \mu_t$ and hence the contribution from $term\,VI$ helps to reduce the critical value of Reynolds number.

### 4. Summary and Conclusions

We have addressed stability characteristics of viscosity stratified shear-thinning fluids in a rotationally-actuated microfluidic platform, considering the Carreau law model for fluid rhgeology. The motivation of the present work stems from the possibility of mixing enhancement and/or seperation in short span through micro channel, which can be useful for medical diagnostics deploying body fluids. The non-vanishing viscosity disturbances are accounted for, which in turn acccomodates for anisotropic perturbations in the stress tensor. Moreover, two different cases, with and without viscosity perturbation are considered. The influence of the rheological parameters, namely, shear-thinning index ($n$) and time constant ($\lambda$), is explored under the circumstance of spanwise rotation.

The results reveal that the dependence of shear-thinning index and time constant on the steady unidirectional base flow profile is less influential, for the time constant value $\lambda < 1.0$. Other two parameters that can strongly affect the stability characteristics of the shear-thinning flow in a rotating channel are the rotation number (which represents the non-dimensional rotation speed of the system) and the Reynolds number (which controls viscous property).

The stability characteristics of the flow are profoundly dependent on the speed of the rotation, and hence the Coriolis force. We have obtained the critical speed of rotation (charactrized by the critical rotation number) for which the onset of transition comes earlier in the case of two-dimensional spanwise disturbances. The critical rotation number for the flow is not constant and decreases as the shear-thinning effect increases. This signifies that even lower rotation speeds may rapidly destabilize the flow within short span, which in effect enhances mixing and/or separation process depending on the application required. Numerical results also confirm that the coupling between viscosity perturbation and rotation itself is very weak for streamwise homogeneous disturbance $(\alpha = 0)$. Primarily, the viscosity distribution and base velocity gradient of shear-thinning flow play important role in the instability mechanism.

The critical value of Reynolds number decreases with increasing $\lambda$ for both with and without viscosity perturbation. Increasing shear-thinning index, in effect, increases the critical Reynolds number for both cases $\mu' \neq 0$ as well as $\mu' = 0$. Interestingly, for the case of time constant $\lambda < 1.0$, a jump in the critical spanwise wave number is observed for $\lambda = 0.75$, implying the flow is more susceptible to rotational instability for shorter wave length. Moreover, the neutral stability boundaries including/excluding viscosity perturbation indicate that, the viscosity perturbation term in the Squire equation is responsible for altering the critical parameters, namely Reynolds number and spanwise wave number of the flow system. For the case of 3D perturbation, stronger rotational effort is required to destabilize the flow as compared to 2D streamwise homogeneous perturbation $(\alpha = 0)$ for $\mu' \neq 0$ and considerably lesser effort is required for the same when viscosity perturbation is not introduced.

The perturbation velocity field is captured to get an overall idea about the instability vortices. The roll cell structures are mostly aligned towards the direction of Coriolis force, and are gradually spread over the entire channel as the time constant $(\lambda)$ increases. The roll-cell pattern forming in the $y - z$ plane shows that the mixing/separation with viscosity perturbation will be more effective with enhanced shear-thinning effects, a consideration of utmost importance towards efficient design of diagnostic devices for rapid biochemical assessment with tiny volumes of body fluid samples.

## Acknowledgement

SC acknowledges Department of Science and Technology, Government of India, for Sir J. C. Bose National Fellowship.

## Declaration of Interest

The authors declare no conflict of interest.